\documentstyle[12pt]{article}

\def\hybrid{\topmargin 0pt      \oddsidemargin 0pt
        \headheight 0pt \headsep 0pt
        \textwidth 6.5in        
        \textheight 9.0in         
        \marginparwidth 0.0in
        \parskip 5pt plus 1pt   \jot = 1.5ex}
\catcode`\@=11
\def\marginnote#1{}

\newcount\hour
\newcount\minute
\newtoks\amorpm
\hour=\time\divide\hour by60
\minute=\time{\multiply\hour by60 \global\advance\minute by-\hour}
\edef\standardtime{{\ifnum\hour<12 \global\amorpm={am}%
        \else\global\amorpm={pm}\advance\hour by-12 \fi
        \ifnum\hour=0 \hour=12 \fi
        \number\hour:\ifnum\minute<10 0\fi\number\minute\the\amorpm}}
\edef\militarytime{\number\hour:\ifnum\minute<10 0\fi\number\minute}

\def\draftlabel#1{{\@bsphack\if@filesw {\let\thepage\relax
   \xdef\@gtempa{\write\@auxout{\string
      \newlabel{#1}{{\@currentlabel}{\thepage}}}}}\@gtempa
   \if@nobreak \ifvmode\nobreak\fi\fi\fi\@esphack}
        \gdef\@eqnlabel{#1}}
\def\@eqnlabel{}
\def\@vacuum{}
\def\draftmarginnote#1{\marginpar{\raggedright\scriptsize\tt#1}}

\def\draft{\oddsidemargin -.5truein
        \def\@oddfoot{\sl preliminary draft \hfil
        \rm\thepage\hfil\sl\today\quad\militarytime}
        \let\@evenfoot\@oddfoot \overfullrule 3pt
        \let\label=\draftlabel
        \let\marginnote=\draftmarginnote
   \def\@eqnnum{(\theequation)\rlap{\kern\marginparsep\tt\@eqnlabel}%
\global\let\@eqnlabel\@vacuum}  }
\newcommand{\l@qq}[2]{\addvspace{2em}
 \hbox to\textwidth{\hspace{1em}\bf #1 \dotfill #2}}

\def\appname{Appendix}
\newcounter{app}
\def\theapp{\Alph{app}}
\def\app{\par
   \addvspace{4ex}
   \@afterindentfalse
  \secdef\@app\@dapp}
\def\@app[#1]#2{\ifnum \c@secnumdepth >\m@ne
        \refstepcounter{app}
        \addcontentsline{toc}{app}{\theapp
        \hspace{1em}#1}\else
      \addcontentsline{toc}{app}{ #1}\fi
   {\parindent \z@ \raggedright
    \Large \bf \appname~\theapp .
   \Large  \bf \hspace{1em}    #2}\nobreak
     \noindent
\setcounter{equation}{0}
\def\theequation{\Alph{app}.\arabic{equation}}}
\def\@dapp#1{%
{\parindent \z@ \raggedright  \bf #1}\par\nobreak}
\def\l@app#1#2{\addpenalty{\@secpenalty}%
   \addvspace{1em plus\p@}%
   \begingroup
   \@tempdima 3em
     \parindent \z@ \rightskip \@pnumwidth
     \parfillskip -\@pnumwidth
     { \bf
     \leavevmode
     #1\hfil \hbox to\@pnumwidth{\hss #2}}\par
     \nobreak
   \endgroup}
\newcounter{sapp}[app]
\def\thesapp{\Alph{app}.\arabic{sapp}}
\def\sapp{\par
   \@afterindentfalse
  \secdef\@sapp\@dsapp}
\def\@sapp[#1]#2{\ifnum \c@secnumdepth >\m@ne
        \refstepcounter{sapp}
        \addcontentsline{toc}{sapp}{\thesapp
        \hspace{1em}#1}\else
      \addcontentsline{toc}{sapp}{ #1}\fi
   {\parindent \z@ \raggedright
    \large \bf \thesapp
   \large  \bf \hspace{1em}    #2}\nobreak
   \vskip 4ex   \noindent
\def\theequation{\Alph{app}.\arabic{equation}}}
\def\@dsapp#1{%
{\parindent \z@ \raggedright  \bf #1}\par\nobreak}
\def\l@sapp#1#2{\addpenalty{\@secpenalty}%
   \begingroup
   \@tempdima 3em
     \parindent \z@ \rightskip \@pnumwidth
     \parfillskip -\@pnumwidth
     { \hspace{1em}
     \leavevmode
     #1 \hfil \dotfill \hbox to\@pnumwidth{\hss #2}}\par \nobreak
     \endgroup}
\def\titlepage{\@restonecolfalse\if@twocolumn\@restonecoltrue\onecolumn
     \else \newpage \fi \thispagestyle{empty}\c@page\z@
        \def\thefootnote{\fnsymbol{footnote}} }

\def\endtitlepage{\if@restonecol\twocolumn \else  \fi
        \def\thefootnote{\arabic{footnote}}
        \setcounter{footnote}{0}}  
\relax

\hybrid

\makeatletter
\newdimen\normalarrayskip              
\newdimen\minarrayskip                 
\normalarrayskip\baselineskip
\minarrayskip\jot
\newif\ifold             \oldtrue            \def\new{\oldfalse}
\def\arraymode{\ifold\relax\else\displaystyle\fi} 
\def\eqnumphantom{\phantom{(\theequation)}}     
\def\@arrayskip{\ifold\baselineskip\z@\lineskip\z@
     \else
     \baselineskip\minarrayskip\lineskip2\minarrayskip\fi}

\def\@arrayclassz{\ifcase \@lastchclass \@acolampacol \or
\@ampacol \or \or \or \@addamp \or
   \@acolampacol \or \@firstampfalse \@acol \fi
\edef\@preamble{\@preamble
  \ifcase \@chnum
     \hfil$\relax\arraymode\@sharp$\hfil
     \or $\relax\arraymode\@sharp$\hfil
     \or \hfil$\relax\arraymode\@sharp$\fi}}

\def\@array[#1]#2{\setbox\@arstrutbox=\hbox{\vrule
     height\arraystretch \ht\strutbox
     depth\arraystretch \dp\strutbox
     width\z@}\@mkpream{#2}\edef\@preamble{\halign \noexpand\@halignto
\bgroup \tabskip\z@ \@arstrut \@preamble \tabskip\z@ \cr}%
\let\@startpbox\@@startpbox \let\@endpbox\@@endpbox
  \if #1t\vtop \else \if#1b\vbox \else \vcenter \fi\fi
  \bgroup \let\par\relax
  \let\@sharp##\let\protect\relax
  \@arrayskip\@preamble}
\def\eqnarray{\stepcounter{equation}%
              \let\@currentlabel=\theequation
              \global\@eqnswtrue
              \global\@eqcnt\z@
              \tabskip\@centering
              \let\\=\@eqncr
              $$%
 \halign to \displaywidth\bgroup
    \eqnumphantom\@eqnsel\hskip\@centering
    $\displaystyle \tabskip\z@ {##}$%
    &\global\@eqcnt\@ne \hskip 2\arraycolsep
         $\displaystyle\arraymode{##}$\hfil
    &\global\@eqcnt\tw@ \hskip 2\arraycolsep
         $\displaystyle\tabskip\z@{##}$\hfil
         \tabskip\@centering
    &{##}\tabskip\z@\cr}
\makeatother

\font\numbers=cmss12
\font\upright=cmu10 scaled\magstep1
\def\stroke{\vrule height8pt width0.4pt depth-0.1pt}
\def\topfleck{\vrule height8pt width0.5pt depth-5.9pt}
\def\botfleck{\vrule height2pt width0.5pt depth0.1pt}
\def\Zmath{\vcenter{\hbox{\numbers\rlap{\rlap{Z}\kern 0.8pt\topfleck}\kern
2.2pt
                   \rlap Z\kern 6pt\botfleck\kern 1pt}}}
\def\Qmath{\vcenter{\hbox{\upright\rlap{\rlap{Q}\kern
                   3.8pt\stroke}\phantom{Q}}}}
\def\Nmath{\vcenter{\hbox{\upright\rlap{I}\kern 1.7pt N}}}
\def\Cmath{\vcenter{\hbox{\upright\rlap{\rlap{C}\kern
                   3.8pt\stroke}\phantom{C}}}}
\def\Rmath{\vcenter{\hbox{\upright\rlap{I}\kern 1.7pt R}}}
\def\Z{\ifmmode\Zmath\else$\Zmath$\fi}
\def\Q{\ifmmode\Qmath\else$\Qmath$\fi}
\def\N{\ifmmode\Nmath\else$\Nmath$\fi}
\def\C{\ifmmode\Cmath\else$\Cmath$\fi}
\def\R{\ifmmode\Rmath\else$\Rmath$\fi}

\def\bea{\begin{eqnarray}}
\def\eea{\end{eqnarray}}
\def\nn{\nonumber}

\def\beq{\begin{equation}}
\def\eeq{\end{equation}}
\def\be{\beq\new\begin{array}{c}}
\def\ee{\end{array}\eeq}

\def\stackreb#1#2{\mathrel{\mathop{#2}\limits_{#1}}}

\def\res{{\rm res}}

\def\d{\partial}

\def\2{{1\over 2}}
\def\N2{${\cal N}=2$}
\def\1N{${\cal N}=1$}

\begin{document}


\begin{titlepage}
\setcounter{footnote}1
\begin{center}
\hfill FIAN/TD-04/98\\
\hfill ITEP/TH-51/98\\
\hfill hep-th/9809196\\
\vspace{0.3in}
{\Large\bf  Seiberg-Witten Systems and Whitham Hierarchies:}\\
\vspace{0.3cm}
{\Large\bf a Short Review}
\footnote{Based on talks given at
"Quarks-98", Suzdal, Russia, 18-24 May 1998.}
\\
\bigskip
\bigskip
{\large A.Marshakov
\footnote{E-mail address: mars@lpi.ac.ru, andrei@heron.itep.ru,
marshakov@nbivms.nbi.dk}
and A.Mironov\footnote{E-mail address: mironov@lpi.ac.ru,
mironov@heron.itep.ru}}
\\
\bigskip
{\it Theory Department,  P.N.Lebedev Physics
Institute, Moscow,~117924, Russia\\
and ITEP, Moscow 117259, Russia}
\end{center}
\bigskip \bigskip

\begin{abstract}
We briefly review the Whitham hierarchies and their applications to
integrable systems of the Seiberg-Witten type. The simplest
example of the $N=2$ supersymmetric $SU(2)$ pure gauge theory is
considered in detail and the corresponding Whitham solutions are
found explicitely.
\end{abstract}

\end{titlepage}

\setcounter{footnote}0

\section{Introduction}

Since constructing by N.Seiberg and E.Witten the low-energy exact solution to
the $N=2$ supersymmetric $SU(2)$ pure gauge theory \cite{SW1}, its
intensive studies have been performed from many different points of view.
In particular, it was realized that this solution as well as analogous
solutions in many other similar theories ($N=2$ supersymmetric gauge theories
in $4d$,
$5d$ and $6d$ with different gauge groups and different matter contents) are
effectively expressed in terms of integrable systems
\cite{GKMMM,int_gen,int_hd,ItoMor,int_rev}. These underlying integrable
systems encode symmetry properties of the solutions.

Moreover, how it was clear from the very beginning \cite{GKMMM}, the
Seiberg-Witten solutions have something to do with {\it two} different
structures: with finite-gap solutions of integrable systems of the
Hitchin type or spin chains (so that the integrals of motion are
fixed and related to the vacuum expectation values
(vev's) of the scalar fields in the supersymmetric theories)
{\it and} with the corresponding {\it Whitham} hierarchy constructed in the
vicinity of  these
solutions and describing the evolution of integrals of motion.
Although the role of both these structures is not quite
understood yet, one can indicate associated key quantities
that play the crucial role for the Seiberg-Witten construction.

The first object is prepotential which defines the low-energy
effective action of the vector multiplets in \N2 supersymmetric theories.
The notion of prepotential can be
easily introduced for the finite-gap solutions: it is a function of integrals
of motion (or of moduli of the complex curve describing the solution) whose
second derivatives give the period matrix. As for the Whitham hierarchy, the
prepotential plays there the role of the (quasiclassical) $\tau$-function.
However, it turns out that one can introduce even more -- the prepotential
which depends on both the Whitham times and the integrals of motion as on
{\it independent} variables (see, e.g.  \cite{RG}). Such prepotential
incorporates properties of the both systems and encodes non-trivial
quantities (correlators) in the physical (supersymmetric) theories (see, e.g.
\cite{Los}, where these correlators have been treated within the framework of
the Donaldson topological theories).

In this note, we are going to review the properties of
prepotential, and corresponding integrable structures arising in the
context of the Seiberg-Witten solution with main emphasis on the Whitham
type hierarchies. As our main illustrative example, we shall consider
the supersymmetric $SU(2)$ pure gauge theory.

First of all, we
consider the finite-gap solution to the integrable hierarchy underlying the
Seiberg-Witten system (which is the periodic Toda chain \cite{GKMMM}). This
solution is described by complex (spectral) curve whose moduli are the
integrals of motion. The symplectic structure can be
defined via meromorphic differential on complex curve and the periods
of this differential define the prepotential.

Then, we introduce evolution of moduli satisfying the equations of the
Whitham hierarchy. After this, we find a generic solution to
the Whitham hierarchy associated with the Seiberg-Witten anzatz and define
the prepotential that depends on both Whitham times {\it and} moduli of
the curve as on independent variables. We illustrate our consideration in the
simplest toric example.

\section{Seiberg-Witten integrable system}

We are going to discuss here the simplest Seiberg-Witten solution
describing the low-energy limit of
the pure gauge theory with $SU(N)$ gauge group. The underlying integrable
is the Toda chain of length $N$ with periodic boundary conditions
\cite{GKMMM}.
In this case, the spectral curve has the form
\be
w + {\Lambda^{2N}\over w} = P_N(\lambda )
\label{suncu}
\ee
or, in hyperelliptic parametrization,
\be\label{2}
y^2 = P_N(\lambda )^2 - 4\Lambda ^{2N}
\ee
where $P_N(\lambda)\equiv\lambda^N-\sum_{k=2}^Nu_k\lambda^{N-k}$ is a
polynomial of degree $N$ so that the genus of the curve is $N-1$, $\Lambda$
is a scale parameter of the physical theory, which can be
easily put equal to unity, $y=w - {\Lambda^{2N}\over w}$. All the dependence
on moduli (vev's of the scalars in the physical theory) is hidden in
coefficients $u_k$'s of $P_N(\lambda)$.  Note that a finite-gap solution to
integrable system usually corresponds to a spectral curve with
{\it marked points}. In
the particular case of Toda chain, these are two points $w=0$ and
$w=\infty$ (and $\lambda=\infty_\pm$ on the two $\lambda $-sheets).

To define the prepotential, one also needs a generating differential
that can be constructed using the symplectic form $\Omega$ of integrable
system. The result for the Toda chain differential is
\be
dS_{SW} = ``\delta^{-1}\Omega\hbox{''} = \lambda dz
\ee
the precise meaning of $``\delta^{-1}\Omega$'' being explained
along with the origin of the whole formula, e.g. in
\cite{KriPh}. Here
\be
dz \equiv d\log w = {dw\over w} = {dP_N\over y}
\ee
The important property of the, generally meromorphic (in the marked points),
differential $dS_{SW}$ is that its derivatives w.r.t. moduli are {\it
holomorphic} differentials. Now the prepotential can be defined by
\be\label{ttdual}
a^D_i = {\d{\cal F}\over\d a_i}
\ee
where\footnote{
We define the symbols $\oint$ and $res$ with
additional factors $(2\pi i)^{-1}$ so that
$$
{\res}_0 \frac{d\xi}{\xi}
= -{\res}_\infty \frac{d\xi}{\xi} =
\oint \frac{d\xi}{\xi} = 1
$$
This explains the appearance of $2\pi i$ factors in the
Riemann identities (like eq.(\ref{pro}) below) and thus in all
definitions of sect.5.
Accordingly, the theta-functions are periodic
with period $2\pi i$, and
$$
\frac{\partial\theta(\vec\xi|{\cal T})}{\partial{\cal T}_{ij}}
= i\pi\partial^2_{ij}\theta(\vec\xi|{\cal T})
$$
since periods of
the Jacobi transformation $\xi_i \equiv \int^\xi d\omega_i$
belong to $2\pi i \left( \Z + {\cal T}\Z\right)$.
}
\be\label{periods}
a_i = \oint_{A_i}dS_{SW}
\ \ \ \ \ \ \
a^D_i = \oint_{B_i}dS_{SW}
\ee
with $A_i\circ B_j=\delta_{ij}$
being canonically normalized cycles on the complex curve.
Note that
\be
{\d dS_{SW}\over\d a_i}=d\omega_i,\ \ \ \
{\d^2{\cal F}\over \d a_i\d a_j}={\d a^D_i\over\d a_j}={\cal T}_{ij}
\ee
where $d\omega_i$ are canonical holomorphic differentials and
${\cal T}_{ij}$ is
the period matrix of the complex curve.

\paragraph{$SU(2)$ example.}
In the simplest $SU(2)$ case, $P_2(\lambda)=\lambda^2 -u$,
the curve (\ref{suncu}) turns into
\be\label{su2}
2{\Lambda ^2}\cosh z = \lambda ^2 - u
\ee
and coincides with the equation relating constant Hamiltonian (energy) with
the co-ordinate $z=iq$ ($z=q$) and momentum $\lambda = p$ of a particle moving
into the
``1-dimensional'' sine-Gordon (sinh-Gordon) -- or
2-site Toda-chain potential:
\be\label{su2em}
2{\Lambda ^2}\cos q = p^2 - u\ \ \ \
\left(2{\Lambda ^2}\cosh q = p^2 - u\right)
\ee
The Toda system as any
1-dimensional system with conserved energy is integrable (since one
only needs a single integral of motion).
One can simply perform the integration of (\ref{su2em})
\be\label{holsu2}
dt = {dq\over p} = {dz\over\lambda} = 2{d\lambda\over y}
\ee

\section{Generalities on Whitham}

Now we turn to general description of the Whitham hierarchy.
Following I.Krichever \cite{Kri,KriW}, we consider
initially a {\it local} system
of holomorphic functions $ \Omega _I$ on complex curve, i.e. functions of
some local parameter $\xi$ in a neighborhood of a point $P$.
One can introduce a set of parameters $t_I$ and define on the space
$(\xi, t_I)$ a $ 1$-form
\begin{equation}\label{1form}
\omega = \sum_I{\Omega _I(P,{\bf t}) \delta t_I}
\end{equation}
Considering its total external derivative
$\delta \omega = \sum_I\delta {\Omega _I(P,{\bf t}) \wedge\delta t_I}$, where
$\delta \omega
= \partial _{\xi} \Omega _I \delta\xi\wedge\delta t_I +
\partial _J \Omega _I\delta t_I \wedge\delta t_J$, one can define general
{\it Whitham equations} as
\begin{equation}\label{gwhi}
\delta \omega\wedge\delta \omega = 0
\end{equation}
so that it is necessary to check the independent vanishing of two
different terms: $ \delta t ^4$ and $ \delta t^3\delta\xi$.
The second term gives
\begin{equation}\label{3whi}
\sum{  \partial _\xi \Omega _{[I}\partial _J \Omega _{C]}} = 0
\end{equation}
where $[...]$ means antisymmetrization. Fixing now some $I=I_0$ and
introducing the ``Darboux'' co-ordinates
\be\label{t0}
t_{I_0}\equiv x \ \ \ \ \ \ \ \Omega _{I_0}(P,{\bf t})\equiv p
\ee
one can get from (\ref{3whi}) the Whitham equations in their standard
form \cite{KriW}
\be\label{pwhi}
\partial _I \Omega _J - \partial _J \Omega _I + \left\{ \Omega _I ,
\Omega _J \right\} = 0
\\
\left\{ \Omega _I , \Omega _J \right\} \equiv {\partial \Omega _I \over
\partial x}{\partial \Omega _J \over \partial p} - {\partial \Omega
_J \over \partial x}{\partial \Omega _I \over \partial p}
\ee
The explicit form of the equations (\ref{pwhi}) strongly depends on choice
of the local co-ordinate $p$. Equations (\ref{gwhi}), (\ref{3whi}) and
(\ref{pwhi}) are defined only locally and have a huge amount of solutions.

Now let us note that the Whitham equations (\ref{pwhi}) can be considered as
compatibility conditions of the system
\be\label{lam-coo}
{\d\lambda\over\d t_I} = \{\lambda ,\Omega _I\}
\ee
The function $\lambda$ can be used itself in order to define a new local
parameter. In this co-ordinate,
the system (\ref{pwhi}) turns into \cite{KriW}
\be\label{lam-whi}
\left.{\d\Omega _I(\lambda,t)\over\d t_J}\right|_{\lambda=const}
= \left.{\d\Omega _J(\lambda,t)\over\d t_I}\right|_{\lambda=const}
\ee
or for the differentials $d\Omega_I \equiv d_{\Sigma}\Omega_I =
{\d\Omega_I\over\d\lambda}d\lambda$
\be\label{lam-whid}
\left.{\d d\Omega _I(\lambda,t)\over\d t_J}\right|_{\lambda=const}
= \left.{\d d\Omega _J(\lambda,t)\over\d t_I}\right|_{\lambda=const}
\ee
The Whitham equations in the form (\ref{lam-whi}) imply the existence
of potential $S$ such that
\be\label{potential}
\left.\Omega _I(\lambda,t)\right|_{\lambda=const}
= \left.{\d S(\lambda,t)\over\d t_I}\right|_{\lambda=const}
\ee
With this potential, the 1-form $\omega$ (\ref{1form}) can be rewritten as
\be
\omega=\delta S - {\d S(\lambda,t)\over\d\lambda}d\lambda
\ee
To fill all these formulas by some real content,
one can consider interesting and important
examples of solutions to the Whitham hierarchy arising when one
takes the basis functions $\Omega _I$ to be the ``quasi-energies'' of the
finite-gap solution;
these are the (globally multivalued) functions
whose periods determine the ``phase'' of the quasi-periodic Baker-Akhiezer
function corresponding to this finite-gap solution.
The potential should be then identified with
\be\label{intds}
S = \int^P dS
\\
{\d S\over\d t_I} = \int^P {\d dS\over\d t_I} = \int^P d\Omega _I =
\Omega _I
\ee
where $dS$ is a generating differential for the finite gap system
\cite{KriPh}.
Usually for KP/Toda hierarchy these functions are taken to be
``half''-multivalued, i.e. their ${\bf A}$-periods are fixed to be zero:
$\oint_{\bf A}d\Omega_I = 0$. The equations (\ref{intds}) are obviously
satisfied in a trivial way in the case of the (finite-gap solutions to)
KP/Toda
theory, when $\{ t_I\}$ are taken to be {\em external} parameters and
$\{\Omega _I\}$ do not depend themselves on $\{ t_I\}$. The idea, however, is
to deform this trivial solution to (\ref{gwhi}), (\ref{3whi}) and
(\ref{pwhi}) into a nontrivial one by {\em a flow} in the moduli space.
Practically it means that the formulas (\ref{intds}) should be preserved even
when $\{\Omega _I\}$ {\em depend} on $\{ t_I\}$, but in a special way --
determined by the equations of Whitham hierarchy. In other words, the Whitham
equations correspond to the dynamics on moduli space of complex curves and
the moduli become depending on the Whitham times.

To realize all this technically, one choose $d\Omega_I$ to be a set of
(normalized) differentials on the complex curve holomorphic outside the
marked points where they have some fixed behavior.
%
An important example of the $\lambda $ co-ordinate is the
{\em hyperelliptic} co-ordinate on hyperelliptic curve
\be
y^2= R(\lambda) = \prod_{\beta}
^{2g+2} (\lambda - r_{\beta})
\ee
Properly choosing the differentials $d\Omega_I$ and
expanding the equations
(\ref{lam-whid}) near the point $\lambda = r_{\alpha}$ one gets
the Whitham equations over finite-gap solutions
\be
{\d r_{\alpha}\over\d t_I} =
\left.{d\Omega_J\over d\Omega_I}\right|_{\lambda=r_{\alpha}}
{\d r_{\alpha}\over\d t_J} \equiv
v_{IJ}^{(\alpha )}({\bf r}){\d r_{\alpha}\over\d t_J}
\label{whirie}
\ee
The set of ramification points $\{ {\bf r}\} = \{ r_{\beta}\}$ is usually
called Riemann invariants for the Whitham equations.

The particular case of Whitham hierarchy associated with the finite-gap
solution of the KdV hierarchy described by the hyperelliptic curve
\be\label{KdVsc}
y^2= R(\lambda) = \prod_{\beta}
^{2g+1} (\lambda - r_{\beta})
\ee
was considered first in \cite{FM}.

Note also that in the framework of Whitham hierarchy
it becomes possible to introduce an analog of the tau-function,
again a generating function for the solutions, which can be symbolically
defined as \cite{KriW}
\be\label{taukri}
\log{\cal T}_{\rm Whitham} = \int_{\Sigma}\bar dS\wedge dS
\ee
where the two-dimensional integral over $\Sigma$ is actually ``localized''
on the non-analyticities of $S$.

\section{Whitham hierarchy for the finite-gap solutions}

Now let us discuss how one can choose the differentials
$d\Omega_I$ in order to get non-trivial solutions to Whitham hierarchy.
They can be defined for the following set of data:
\begin{itemize}
\item  complex curve (Riemann surface) of genus $g$;
\item  a set of punctures $P_i$ (marked points);
\item co-ordinates $\xi_i$ in the vicinities of the punctures $P_i$;
\item pair of differentials, say, ($d\lambda$, $dz$) with fixed periods.
\end{itemize}

We start with considering the simplest case of a single
puncture, say at $\xi = \xi_0 = 0$. This situation is typical, e.g., for
the Whitham hierarchy in the vicinity of the finite-gap solutions to
KP or KdV hierarchies. For instance, in the KdV case the
spectral curve has the form (\ref{KdVsc}) so that the marked point is at
$\lambda=\infty$ but, in contrast to the Toda curve (\ref{2}), there is only
one infinite point (since it is the ramification point).

Given this set of data, one can introduce meromorphic
differentials with the poles only at some point $P_0$
such that in some local co-ordinate $\xi$: $\xi (P_0)=\xi_0=0$
\be
d\Omega_n \stackreb{P\to P_0}{=} \left( \xi^{-n-1} + O(1) \right)d\xi,
\ \ \  n\geq 1
\label{canbas}
\ee
This condition defines $d\Omega_n$ up to arbitrary
linear combination of $g$ holomorphic differentials
$d\omega_i$, $i = 1,\ldots,g$ and there are two different
natural ways to fix this ambiguity.
The first way already mentioned
above is to require that $d\Omega_n$ have vanishing
$A$-periods,
\be
\oint_{A_i} d\Omega_n = 0    \ \ \ \forall i,n
\label{vanAp}
\ee
Their generating functional ($\zeta\equiv\xi (P')$, $\xi\equiv\xi (P)$)
\be
W(\xi,\zeta) = \sum_{n=1}^\infty n\zeta^{n-1} d\zeta d\Omega_n(\xi)
+ \dots
\ee
is well known in the theory of Riemann surfaces.
It can be expressed through the Prime form $E(P,P')$ (see Appendix A
of \cite{RG} for details):
\footnote{
For example, for genus $g=1$ in the co-ordinate $\xi \sim \xi+1 \sim
\xi+\tau$
the formulas acquire the form
$$
d\Omega_1 = (\wp (\xi) - const)d\xi ,
\ d\Omega_2 = -\frac{1}{2}\wp'( \xi )d \xi ,
\ \ldots, d\Omega_n = \frac{(-)^{n+1}}{n!}\partial^{n-1}\wp(\xi)d\xi
$$
and
$$
W(\xi,\xi') = \sum_{n=1}^\infty \frac{(-)^{n+1}\xi'^{n-1}}{(n-1)!}
\frac{\partial^{n-1}}{\partial \xi ^{n-1}}\wp( \xi )d \xi d\xi'
- const\cdot d \xi  d\xi' =
\wp(\xi  - \xi')d\xi  d\xi' =
\partial_\xi  \partial_{\xi'} \log \theta_*(\xi - \xi')
$$
where $*$ denotes the (on torus the only one) odd theta-characteristic.
For $g=1$ $\nu_*^2(\xi)
=\theta_{*,i}(0)d\omega_i$ is just $d\xi$. Let us also point out that
choosen in this way
co-ordinate $\xi$ is not convenient from the point of view of Whitham
hierarchy since its "periods" ($\tau = \oint_B d\xi$) depend on moduli
of the curve.}
\be
W(P,P') = \partial_P \partial_{P'} \log E(P,P')
\label{WE}
\ee
Such  $W(P,P')$ has a second order pole on diagonal $P\to P'$,
\be
W(\xi,\zeta) \sim \frac{d\xi d\zeta}{(\xi - \zeta)^2} + O(1) =
\sum_{n=1}^\infty n\frac{d\xi}{\xi^{n+1}}\zeta^{n-1}d\zeta + O(1)
\label{Wexp}
\ee
It is
the differentials (\ref{canbas}) that should be related with the potential
differential $dS$ by
\be
d\Omega_n = {\d dS\over\d t_n}
\ee
The second way to normalize differentials is to impose the condition
\be
\frac{\partial d\hat\Omega_n}{\partial\ {\rm moduli}} =
{\rm holomorphic}
\label{derdS1}
\ee
so that these differentials becomes similar to the generating differential
$dS_{SW}$ of integrable system (see sect.2).

Now let us specialize this description to the case considered in
sect.2, i.e. to the family of spectral curves
(\ref{suncu}). The first problem is that naively there are no
solutions to eq.(\ref{derdS1}) because the curves (\ref{suncu}) are
spectral curves of the {\it Toda-chain} hierarchy (not of KP/KdV type).  The
difference is that adequate description in the Toda case requires {\it
two} punctures instead of one. As already mentioned in sect.2 above, the
curves (\ref{suncu}) have two marked points and
there exists a function $w$
with the $N$-degree pole and zero at two (marked) points $\lambda =
\infty_\pm$, where $\pm$ labels two sheets of the hyperelliptic
representation (\ref{suncu}), $w(\lambda = \infty_+) = \infty$, $w(\lambda =
\infty_-) = 0$.  Accordingly, there are two families of the differentials
$d\Omega_n$:  $d\Omega^+_n$ with the poles at $\infty_+$ and
$d\Omega^-_n$ with the poles at $\infty_-$.

However, there are {\it no} differentials $d\hat\Omega^\pm$,
only $d\hat\Omega_n = d\hat\Omega^+_n + d\hat\Omega^-_n$, i.e.
condition (\ref{derdS1}) requires $d\hat\Omega_n$ to have the
poles at both punctures. Moreover, the coefficients in front of
$w^{n/N}$ at $\infty_+$ and $w^{-n/N}$ at $\infty_-$
(\ref{canbas}) coincide (in Toda-hierarchy
language, this is the Toda-{\it chain} case with the same dependence upon
negative and positive times).

The differentials $d\hat\Omega_n$ for the family (\ref{suncu})
have the form \cite{RG}:
\be
d\hat\Omega_n = R_n(\lambda)\frac{dw}{w}
= P^{n/N}_+(\lambda)\frac{dw}{w}
\label{expldO}
\ee
The polynomials $R_n(\lambda)$ of degree $n$ in $\lambda$
are defined by the property that
$P'\delta R_n - R_n'\delta P$ is a polynomial of degree
less than $N-1$. Thus, $R_n(\lambda) = P^{n/N}_+(\lambda)$,
where $\left(\sum_{k=-\infty}^{+\infty} c_k\lambda^k\right)_+
= \sum_{k=0}^{+\infty} c_k\lambda^k$.
For example\footnote{In the case of the Toda chain, since there
are two punctures, one can consider the meromorphic differential
with the simple pole at each of them.}:
\be\label{Rn}
R_0=1,\\
R_1 = \lambda, \nn \\
R_2 = \lambda^2 - \frac{2}{N}u_2, \nn \\
R_3 = \lambda^3 - \frac{3}{N}u_2\lambda - \frac{3}{N}u_3, \nn \\
R_4 = \lambda^4 - \frac{4}{N}u_2\lambda^2 - \frac{4}{N}u_3\lambda -
\left(\frac{4}{N}u_4 + \frac{2(N-4)}{N^2}u_2^2\right), \nn \\
\ldots
\ee
These  differentials satisfy (\ref{derdS1})
provided the moduli-derivatives are
taken at constant $w$ (not $\lambda$!). Thus, the
formalism of the previous section is applicable
for the local parameter $\xi = w^{\mp 1/N}$.\footnote{Rule of thumb
generally is to choose as the local co-ordinate for the Whitham hierarchy the
parameter living on the bare spectral curve. For the Toda case,
one may choose as a bare
curve the $w$-cylinder (see, e.g.
\cite{int_rev}), with the corresponding generating differential $dS=\lambda
d\log w$.}

Thus, among the data one needs for the definition of solution to
the Whitham
hierarchy associated with the Toda chain, there are the punctures at
$\lambda = \infty_\pm$ and the relevant co-ordinates in the vicinities of
these punctures $\xi \equiv w^{-1/N} \sim \lambda^{-1}$ at $\infty_+$ and
$\xi \equiv w^{+1/N} \sim \lambda^{-1}$ at $\infty_-$.  The parametrization
in terms of $w$, however, does not allow to use the advantages of
hyperelliptic form (\ref{2}).

Let us consider our simplest explicit example.

\paragraph{Whitham equations for the $SU(2)$ case.}

In the case of 2-site periodic Toda chain with the spectral
curve (\ref{su2}), one can restrict himself to the first two
differentials $d\Omega_n$ and the two first times $t_0$,
$t_1$.
Then, the Whitham equations
\be
{\d d\Omega_i\over\d t_j} = {\d d\Omega_j\over\d t_i}
\label{whitham}
\ee
are reduced to the only equation
\be
{\d d\Omega_0\over\d t_1} = {\d d\Omega_1\over\d t_0}
\label{whithamsl2}
\ee

To write it down explicitely one should remember that
\begin{itemize}
\item
There are two independent differentials (see (\ref{expldO}) and
(\ref{Rn}) and formulas of sect.2)
\be
d\Omega_0 = dz + \gamma_0{d\lambda\over y} =
\left(1 +{\gamma_0\over 2\lambda} \right)dz
\\
d\Omega_1 = \lambda dz + \gamma_1{d\lambda\over y} =
\left(\lambda +{\gamma_1\over 2\lambda} \right)dz
\label{diffw}
\ee
and two corresponding Whitham times $t_0$ and $t_1$.
\item
The coefficients $\gamma_i$ ($i=0,1$) are fixed, as usual, by
vanishing of the correspondent $A$-periods
\be
\oint_A d\Omega_i = 0
\ee
i.e.
\be
\gamma_1 = - {a\over\sigma}
\ \ \ \ \ \ \
\gamma_0 = -{1\over\sigma}
\label{alpha1}
\ee
where
\be
a = \oint_A dS = \oint_A \lambda dz
\\
\sigma = \oint_A {d\lambda\over y} = {\d a\over\d u}
\label{asigma}
\ee
\item
The derivatives over moduli in (\ref{whitham}) are taken at fixed values
of $z$-variable, while the $\lambda$-variable depends on moduli and this
dependence is given by (\ref{su2}). This rule leads to the relations
\be
{\d\lambda\over\d u} = {1\over 2\lambda}
\\
{\d a\over\d u} = \oint_A {\d\lambda\over\d u}dz =
\oint_A{dz\over 2\lambda} = \sigma
\ee
\end{itemize}
Using these relations one easily gets from (\ref{whitham}) the following
Whitham equation on moduli
\be\label{whisl2}
{\d u\over\d t_1} = a(u){\d u\over\d t_0}
\ee
The function $a(u)$ is an elliptic integral. Its explicit expression depends
on choice of the cycles. With the choice used in \cite{RG}, i.e. when
the $\bf A$-cycles encircle the points
$\lambda=\sqrt{u-2\Lambda^2}\equiv r^-$ and
$\lambda=\sqrt{u+2\Lambda^2}\equiv r^+$,
\be\label{elint}
a={2\over\pi}r^+E(k),\ \ \ \ \sigma={1\over\pi r^+}K(k)
\ee
where $K(k)$ and $E(k)$ are complete elliptic integrals of the first and
the second kinds respectively and the elliptic modulus is $k={2\over r^+}$.

The Whitham equation (\ref{whisl2}) has the general solution:
\be\label{sol}
u=F\left(t_0+a(u)t_1\right)
\ee
where $F(x)$ is an arbitrary function. This solution can be also
rewritten in the form
\be\label{soli}
\Phi(u)=t_0+a(u)t_1
\ee
where $\Phi(u)$ is the function inverse to $F(x)$.

\section{Solution to the Whitham hierarchy and prepotential}

Now we are going to extend the system of differentials $d\Omega_i$ introduced
in the previous section to include the holomorphic differentials. This will
allow us to construct the function that, as a function of
one set of variables, is the prepotential of the
Seiberg-Witten theory (as defined in sect.2) and in the other set of {\it
independent} variables is the (logarithm of) $\tau$-function
of the associated solution to
the Whitham hierarchy. On this way, we also get a generic solution to
the associated Whitham equations.

Thus, we extend the set of Whitham equations by the
equations involving the holomorphic differentials $d\omega_i$
and moduli $\alpha_i$:
\be\label{whd}
{\d d\Omega_n\over\d \alpha_i}={\d d\omega_i\over\d t_n},\ \ \ \
{\d d\omega_i\over\d \alpha_j}={\d d\omega_j\over\d \alpha_i}
\ee
This system is solved by the differential $dS$ that satisfies
\be\label{dSe}
\frac{\partial dS}{\partial \alpha_i} = d\omega_i, \ \ \
\frac{\partial dS}{\partial t_n} = d\Omega_n
\label{derfdS}
\ee
Then, the first equation in (\ref{dSe}) implies that $dS$ is to be
looked for as a linear combination of the differentials $d\hat\Omega_n$,
satisfying (\ref{derdS1}).
Let us, following \cite{ItoMor,RG} introduce a
generating functional for $d\hat\Omega_n$
with infinitely many auxiliary parameters $t_n$:
\be
dS = \sum_{n\geq 1} t_n d\hat\Omega_n =
\sum_{i=1}^g \alpha_i d\omega_i + \sum_{n\geq 1} t_nd\Omega_n
\label{dSexp}
\ee
The periods
\be
\alpha_i = \oint_{A_i} dS
\label{perdef}
\ee
can be considered as particular co-ordinates on the moduli space.
Note that these periods do not exactly coincide with (\ref{periods}),
eq.(\ref{perdef}) defines $\alpha_i$ as functions of $h_k$ and $t_n$,
or, alternatively, $h_k$ as functions of $\alpha_i$ and $t_n$
so that derivatives $\partial h_k/\partial t_n$ are non-trivial.
In what follows we shall consider $\alpha_i$ and $t_n$,
\be
t_n = {\res}_{\xi=0}\ \xi^n dS(\xi)
\label{Tdef}
\ee
as independent variables so that partial derivatives w.r.t.
$\alpha_i$ are taken at constant $t_n$ and partial derivatives
w.r.t. $t_n$ are taken at constant $\alpha_i$.

The differential $dS$ (\ref{dSexp}) determines a generic form of the
solution associated to the Seiberg-Witten type Whitham hierarchy. The
Whitham dynamics itself for given $dS$ can be formulated in terms of
equations (\ref{dSe}).  Note that, if one restricts himself to the Whitham
hierarchy with several first times (generally, for the genus $g$ complex
curve there are $g+n-1$ independent times, with $n$ being the number of
punctures), all higher times in (\ref{dSexp}) play the role of constants
(parameters) of generic Whitham solution (see example of $SU(2)$ below).

Note that the Seiberg-Witten differential $dS_{SW}$ is
$dS_{SW} = d\hat\Omega_2$, i.e.
\be
\left.dS\right|_{t_{n}=\delta_{n,1}} = dS_{SW}, \ \ \
\left.\alpha^i\right|_{t_{n}=\delta_{n,1}} = a^i, \ \ \
\left.\alpha^D_i\right|_{t_{n}=\delta_{n,1}} = a^D_i
\ee
and $\alpha$-variables are naturally associated with the Seiberg-Witten
moduli, while $t$-variables -- with the corresponding Whitham times
(although $\alpha_i$'s can be also considered as variables of
Whitham dynamics, c.f. (\ref{whd})).

Now one can introduce the Whitham tau-function (\ref{taukri}) whose logarithm
is a {\it prepotential}
${\cal F}(\alpha_i,t_n)\equiv\log{\cal T}$ by an analog of conditions
(\ref{ttdual})-(\ref{periods}):
\be
\frac{\partial {\cal F}}{\partial \alpha_i} = \oint_{B_i} dS, \ \ \
\frac{\partial {\cal F}}{\partial t_n} = \frac{1}{2\pi in}
 {\res}_0\ \xi^{-n}dS
\label{prepdef}
\ee
Their consistency follows from (\ref{derfdS}) and Riemann
identities. In particular,
\be
\frac{\partial^2{\cal F}}{\partial t_m\partial t_n} =
\frac{1}{2\pi in} {\res}_0\ \xi^{-n} \frac{\partial dS}{\partial t_m}
= \frac{1}{2\pi in}{\res}_0\ \xi^{-n} d\Omega_m =
\frac{1}{2\pi im} {\res}_0\ \xi^{-m} d\Omega_n
\label{secderF}
\ee
From this calculation, it is clear that the definition
(\ref{prepdef}) assumes that co-ordinates $\xi$ are not changed
under the variation of moduli. It means that they provide a
moduli-independent parametrization of entire family --
like $w$ in the case of (\ref{suncu}). Since
moduli-independence of $\xi$ should be also consistent with
(\ref{derdS1}), the choice of $\xi$ is strongly restricted:
to $w^{\pm 1/N}$ in the case of (\ref{suncu}) (see the discussion in the
previous section).

The last relation in (\ref{secderF})
(symmetricity) is just an example of the Riemann relations and it
is proved by the standard argument:
\be
0 = \int d\Omega_m \wedge d\Omega_n =
\oint_{A_i} d\Omega_m \oint_{B_i} d\Omega_n -
\oint_{A_i} d\Omega_n \oint_{B_i} d\Omega_m  +
\frac{1}{2\pi i}{\res} \left(d\Omega_m d^{-1} d\Omega_n\right) =\\=
0 + 0 +  \frac{1}{2\pi in}{\res}_0\ \xi^{-n} d\Omega_m -
\frac{1}{2\pi im}{\res}_0\ \xi^{-m} d\Omega_n
\label{pro}
\ee
where (\ref{vanAp}) and (\ref{canbas}) are used at the final
stage. The factors like $n^{-1}$ arise since
$\xi^{-n-1}d\xi = -d(\xi^{-n}/n)$.  We shall also
use a slightly different normalization $d\Omega_n \sim
w^{n/N}\frac{dw}{w} = \frac{N}{n}dw^{n/N}$, accordingly the residues
in (\ref{pro}) and (\ref{prepdef}) and (\ref{secderF})
will be multiplied by $N/n$ instead of $1/n$.

By definition, the prepotential is a homogeneous function of
its arguments $a_i$ and $t_n$ of degree 2,
\be
2{\cal F} =
\alpha_i \frac{\partial{\cal F}}{\partial \alpha_i} +
t_n\frac{\partial{\cal F}}{\partial t_n} =
\alpha_i\alpha_j
\frac{\partial^2{\cal F}}{\partial \alpha_i\partial \alpha_j} +
2\alpha_i t_n
\frac{\partial^2{\cal F}}{\partial \alpha_i\partial t_n} +
t_mt_n \frac{\partial^2{\cal F}}{\partial t_m\partial t_n}
\label{hom}
\ee
Again, this condition can be
proved with the help of Riemann identities, starting from
(\ref{prepdef}), (\ref{perdef}) and (\ref{Tdef}).
At the same time, ${\cal F}$ is not just a quadratic
function of $a_i$ and $t_n$, a non-trivial dependence on
these variables arise through the dependence of $d\omega_i$
and $d\Omega_n$ on moduli (like $u_k$ or $h_k$)
which in their turn depend on $a_i$ and $t_n$. This
dependence is obtained, for example, by substitution of (\ref{dSexp})
into (\ref{derfdS}):
\be\label{whian}
d\hat\Omega_n + t_m\frac{\partial d\hat\Omega_m}{\partial u_l}
\frac{\partial u_l}{\partial t_n} = d\Omega_n,
\ee
i.e.
\be
\left(\sum_{m,l} t_m\frac{\partial u_l}{\partial t_n}\right)
\oint_{A_i}\frac{\partial d\hat\Omega_m}{\partial u_l}
= - \oint_{A_i} d\hat\Omega_n
\ee
The integral in the l.h.s. is expressed, according to (\ref{derdS1}),
through the integrals of holomorphic 1-differentials, while
the integral in the r.h.s. --  through the periods of $dS$.
If
\be
\oint_{A_i} \frac{\partial dS}{\partial u_l} =
\oint_{A_i} dV_l = \Sigma_{il}
\ee
then
\be
t_m\frac{\partial u_k}{\partial t_m}
= \Sigma^{-1}_{ki}\alpha_i
\label{Whitheq}
\ee
The relations (\ref{Whitheq}) can be thought of as the other form of the
Whitham hierarchy.

\paragraph{$SU(2)$ case.}

Now let us discuss the solution
(\ref{dSexp}) in the simplest $SU(2)$ case.
First, we consider only two first non-zero times so that $dS$ has a
particular form
\be\label{anzatz}
dS=t_0d\hat\Omega_0+t_1d\hat\Omega_1=t_0dz+t_1\lambda dz
\ee
Then,
using formulas of the end of the previous section for the
differentials $d\Omega_{1,2}$, one can easily obtain
from (\ref{whian}) for $n=0,1$ the following two equations:
\be\label{hrena}
t_1{\d u\over\d t_1}=-{a\over\sigma},\\
t_1{\d u\over\d t_0}=-{1\over\sigma}
\ee
Any solution to the equations (\ref{hrena}) evidently solves
simultaneously (\ref{whisl2}), but not {\it vice versa}.

Let us use the manifest form of functions $a(u)$ and $\sigma(u)$ from
(\ref{elint}):
\be
t_1{\d u\over\d t_1}=-{8\over k^2}{E(k)\over K(k)},\\
t_1{\d u\over\d t_0}=-{2\pi\over\ k}{1\over K(k)}
\ee
Then, the first equation of (\ref{hrena}) has the
solution \cite{HB}
\be
t_1=c(t_0){k\over E(k)}
\ee
with arbitrary function $c(t_0)$.
The solution to the both equations (\ref{hrena}) takes the form
\be\label{cf}
a(u)={const-t_0\over t_1}
\ee
This particular solution has to be compared with (\ref{soli}) with the
constant function $\Phi(u)$ (since it has no inverse, the second form
(\ref{sol}) does not exist for this concrete solution).

In order to get the general solution (\ref{soli})
to the Whitham hierarchy, one has to require for
all higher odd times in (\ref{dSexp}) (i.e. with $n>1$)
to be non-zero constants.
Then, these higher times parameterize the general function
$\Phi(u)$. Indeed, how it can be easily obtained from (\ref{whian}),
the solution to the Whitham equations with general $dS$ (\ref{dSexp})
is given by formula (\ref{soli}) with
\be
\Phi'(u)=\sum_{k=1}t_{2k+1}(-)^k{(2k-1)!!\over 2^k}u^k
\ee
Therefore, the differential $dS$ (\ref{dSexp}) actually leads to the general
solution to the Whitham hierarchy.

To complete our consideration of particular
system (\ref{hrena}), let us discuss what happens when one
considers $u$ as
a function of {\it three} independent variables: $t_0$, $t_1$ and
$\alpha$. We are going to demonstrate that, in this case,
we still get the same solution depending on {\it two} independent
variables.

Since
\be\label{alpha}
\alpha\equiv\oint_A dS=t_1a+t_0
\ee
(cf. with eq.(\ref{cf})) and using (\ref{asigma}),
(\ref{hrena})
turns into
\be\label{more}
t_1{\d u\over\d t_1}=-\alpha{\d u\over\d \alpha}+t_0{\d
u\over\d\alpha},\\
{\d u\over\d t_0}=-{\d u\over\d \alpha}
\ee
with the solution
\be\label{solu}
u=\Psi\left({\alpha-t_0\over t_1}\right)
\ee
where $\Psi(x)$ is an arbitrary function. Note that (\ref{cf})
now can be understood just as the condition of constant $\alpha$.

Let us now note that $a(u)$ for the given data is some known
function of $u$. Then, the inverse (not arbitrary!) function $u=\Psi(a)$ can
be rewritten in the form (\ref{solu}) using (\ref{alpha}) with independent
variables $\alpha$, $t_0$ and $t_1$. It means that one can not
consider arbitrary function $\Psi(x)$ with the particular anzatz
of zero higher times in (\ref{dSexp}), i.e.  solutions to the equations
(\ref{more}) should be additionally constrained.

If now one forgets about the specific anzatz for $dS$
with zero higher times and considers
unconstrained $dS$ (\ref{dSexp}), one should
also consider the additonal Whitham equation
\be
{\d d\Omega_n\over\d\alpha} = {\d d\omega\over\d t_n}
\ee
that exactly gives rise to the equations (\ref{more})
\be
{\d u\over\d t_1}=-a{\d u\over\d \alpha},\\
{\d u\over\d t_0}=-{\d u\over\d \alpha}
\ee
These equations are equaivalent to (\ref{whisl2}) with $u$ being a function
of $t_0-\alpha$. It means that we again obtain the same solution
(\ref{soli}) that describes $u$ as a function of {\it two} independent
variables $t_1$ and $t_0-\alpha$. This perfectly matches the number $g+n-1=2$
of independent variables.

\section{Conclusion}

Thus, we have constructed the prepotential for the Whitham hierarchy
associated with the Seiberg-Witten theory. How it was already explained in
\cite{GKMMM}, the Whitham equations themselves describe the dependence of
effective coupling constants given by the period matrix of the
Seiberg-Witten curve on bare coupling constants $t_n$. The
prepotential also depends on characteristics of the vacuum parametrized by
vev's $u_k$. Using this interpretation, one expects \cite{RG} that the
constructed Whitham prepotential has to be associated with the generating
function of topological correlators, the Whitham times being bare coupling
constants of different local operators\footnote{Note also that, after the
rescaling
\be
h_k \rightarrow t_1^k h_k,\ \ \
{\cal H}_k \rightarrow t_1^k {\cal H}_k
\label{hH}
\ee
$t_1$ can be identified with $\Lambda$ \cite{RG}. Then eqs. (\ref{1der}),
(\ref{2der}) below at $n=1$ are
naturally interpreted in terms of the stress-tensor anomaly,
\be
\ldots +\vartheta^4\Theta^\mu_\mu =
\beta {\rm tr} \Phi^2  = \ldots + \vartheta^4 \beta {\rm tr}
\left(G_{\mu\nu}G^{\mu\nu} + iG_{\mu\nu}\tilde G^{\mu\nu}\right),
\ee
since for any operator ${\cal O}$
\be
\frac{\partial}{\partial \log\Lambda} \langle {\cal O}\rangle \ =
\ \langle \beta {\rm Tr} \Phi^2, {\cal O}\rangle
\label{an}
\ee
Analogous interpretation for $n\geq 2$ involves anomalies of
$W_{n+1}$-structures.
}. It issues the problem of calculating derivatives of the
prepotential with their further comparison with the topological
correlators. Indeed, the first two derivatives have been calculated in
\cite{RG} and turned out to be
\be
\frac{\partial{\cal F}}{\partial t^n} = \frac{\beta}{2\pi i n}
\sum_m  mt_m {\cal H}_{m+1,n+1}
= \frac{\beta}{2\pi in}t_1{\cal H}_{n+1} + O(t_2,t_3,\ldots)
\label{1der}
\ee
\be
\frac{\partial^2{\cal F}}{\partial \alpha^i\partial t^n}
= \frac{\beta}{2\pi in} \frac{\partial {\cal H}_{n+1}}{\partial a^i}
\label{11der}
\ee
\be
\frac{\partial^2{\cal F}}{\partial t^m\partial t^n}
= -\frac{\beta}{2\pi i} \left({\cal H}_{m+1,n+1}
+ \frac{\beta}{mn}\frac{\partial {\cal H}_{m+1}}{\partial a^i}
\frac{\partial {\cal H}_{n+1}}{\partial a^j}
\partial^2_{ij} \log \theta_E(\vec 0|\tau)\right)
\label{2der}
\ee
etc. In these formulas parameter $\beta = 2N$, $m,n = 1,\ldots,N-1$ and
${\cal H}_{m,n}$ are certain homogeneous combinations of
$u_k$, defined in terms of
$P_N(\lambda)$:
\be
{\cal H}_{m+1,n+1} =
-\frac{N}{mn}
{\res}_\infty\left(P^{n/N}(\lambda)d P^{m/N}_+(\lambda)\right)
= {\cal H}_{n+1,m+1}
\ee
and
\be
{\cal H}_{n+1} \equiv {\cal H}_{n+1,2}
= -\frac{N}{n}{\res}_\infty P^{n/N}(\lambda)
d\lambda
\ee

Note that the equations (\ref{1der})-(\ref{2der}) at $n=1$ already appeared
in the context of topological theories \cite{Los}. These are evidently only
the first steps in the direction that presumably should lead us
to understanding of the supersymmetric gauge theories.

\section*{Acknowledgments}
The authors are grateful to H.Braden, A.Gorsky, S.Kharchev, I.Krichever,
A.Morozov and M.Olshanetsky for valuable discussion. We especially acknowledge
the discussions with A.Morozov on Whitham hierarchy. The work of A.Mar. is
partially supported by grants INTAS-96-482 and RFBR-98-01-00344, that of
A.Mir.  -- by grants RFBR-96-02-16210(a), INTAS-1038. A.Mir. also
acknowledges the Royal Society for support under a joint project.


\begin{thebibliography}{12}

\bibitem{SW1}
N.Seiberg and E.Witten, {\sl Nucl.Phys.} {\bf B426} (1994) 19,
hep-th/9407087; {\sl Nucl.Phys.} {\bf B431} (1994) 484,
hep-th/9408099
\bibitem{GKMMM}
A.Gorsky, I.Krichever, A.Marshakov, A.Mironov and A.Morozov,
{\sl Phys.Lett.} {\bf B355} (1995) 466, hep-th/9505035
\bibitem{int_gen}
E.Martinec and N.Warner, {\sl Nucl.Phys.}, {\bf B459} (1996) 97-112;
hep-th/9509161;\\
R.Donagi and E.Witten, {\sl Nucl.Phys.}, {\bf B460} (1996)
299; hep-th/9510101;\\
E.Martinec, {\sl Phys.Lett.}, {\bf B367} (1996) 91-96;
hep-th/9510204;\\
A.Gorsky and A.Marshakov, {\sl Phys.Lett.}, {\bf B375}
(1996) 127, hep-th/9510224;\\
A.Marshakov, {\sl Mod.Phys.Lett.}, {\bf A11}
(1996) 1169; hepth/9602005;\\
A.Gorsky, A.Marshakov, A.Mironov and A.Morozov,
{\sl Phys.Lett.}, {\bf B380}
(1996) 75; hep-th/9603140;\\
I.Krichever and D.Phong, {\sl J.Diff.Geom.}, {\bf 45} (1997)
349-389; hep-th/9604199;\\
A.Gorsky, S.Gukov and A.Mironov, {\sl Nucl.Phys.}, {\bf B517} (1998)
409-461; hep-th/9707120;\\
E.D'Hoker and D.Phong, hepth/9709053, hepth/9804126
\bibitem{int_hd}
N.Nekrasov, hepth/9609219;\\
A.Marshakov, M.Martellini and A.Morozov,
{\sl Phys.Lett.}, {\bf B418} (1998) 294;\\
A.Gorsky, S.Gukov and A.Mironov, {\sl Nucl.Phys.}, {\bf B518} (1998)
689-713; hep-th/9710239;\\
A.Marshakov, hepth/9709001;\\
A.Marshakov and A.Mironov, {\sl Nucl.Phys.}, {\bf B518} (1998) 59-91;
hep-th/9711156
\bibitem{ItoMor}
H.Itoyama and A.Morozov,
{\sl Nucl.Phys.}, {\bf B491} (1997) 529, hep-th/9512161
\bibitem{int_rev}
A.Marshakov, {\sl Theor.Math.Phys.}, 112 (1997) 791, hep-th/9702083;\\
A.Mironov, hep-th/9801149;\\
and references in these reviews
\bibitem{RG}
A.Gorsky, A.Marshakov, A.Mironov and A.Morozov, {\sl Nucl.Phys.}, {\bf B527}
(1998) 690-716; hep-th/9802007
\bibitem{Los}
G.Moore and E.Witten, hep-th/9709193;\\
A.Losev, N.Nekrasov and S.Shatashvili, hep-th/9711108; hep-th/9801061;\\
G.Moore and M.Mari\~no, {\sl Nucl.Phys.Proc.Suppl.}, {\bf 68}
(1998) 336-347; hep-th/9712062
\bibitem{KriPh}
B.Dubrovin, I.Krichever and S.Novikov, in {\sl Sovremennye problemy
matematiki}, VINITI, 1984;\\
A.Marshakov, hepth/9607159;\\
I.Krichever and D.Phong, hep-th/9708170 and references therein
\bibitem{SW3}
N.Seiberg and E.Witten, hep-th/9607163
\bibitem{Kri} I.Krichever, {\sl Uspekhi Mat.Nauk}, {\bf 44:2} (1989) 121
\bibitem{KriW}
I.Krichever, hep-th/9205110; {\sl Comm.Math.Phys.} {\bf 143}
(1992) 415
\bibitem{FM}
H.Flashka, G.Forest and D.W.McLaughlin, {\sl Comm.Pure Appl.Math.},
{\bf 33} (1979) 299-329
\bibitem{HB} H.Braden, A.Mironov and M.Olshanetsky, in progress

\end{thebibliography}
\end{document}